# Rejuvenation of metallic glasses under high pressure

C. Wang,[1] Z. Z. Yang,[1] T. Ma,[1,2] Y. T. Sun,[1] Y. Y. Yin,[1] Y. Gong,[1] L. Gu,[1,3] P. Wen,[2] P. W. Zhu,[1] Y. W. Long,[1] X. H. Yu,[1,*] C. Q. Jin,[1] W. H. Wang,[1] H. Y. Bai[1*]

[1]*Beijing National Laboratory for Condensed Matter Physics and Institute of Physics, Chinese Academy of Sciences, Beijing 100190, China.*

[2]*State Key Laboratory of Superhard Materials, College of Physics, Jilin University, Changchun 130012, China.*

[3] *Collaborative Innovation Center of Quantum Matter, Beijing 100190, China.*

**Abstract**

Modulating energy states of metallic glasses (MGs) is significant in understanding the nature of glasses and control their properties. In this study, we show that rejuvenation in enthalpy can be achieved and preserved in bulk MGs by using high pressure (HP) annealing, which is a controllable method to continuously alter the energy states of MGs. Contrary to the decrease in enthalpy by conventional annealing at ambient pressure, such rejuvenation can occur and be enhanced by increasing both of annealing temperature and pressure. By using double aberration corrected scanning transmission electron microscopy, it is revealed that the rejuvenation, which is attributed to coupling effect of high pressure and high temperature, originates from the microstructural change that involves "negative flow units" with a higher atomic packing density compared to that of the elastic matrix of MGs. The results demonstrate that HP-annealing is an effective way to rejuvenate MGs into higher energy states, and it may assist in understanding the microstructural origin of the rejuvenation in MGs.

*Corresponding authors: hybai@iphy.ac.cn; yuxh@iphy.ac.cn

Metallic glasses (MGs) are quenched from melts and inherit the intrinsic topological and geometrical microstructural frustrations with substantial quenched-in "defects"[1,2]. Confirmed by both experiments and simulations[3,4], MGs display microstructural heterogeneity, characterized by the closely packed elastic matrix and loosely packed atomic regions in nanoscale[5]. The structures, size and distribution of the closely and loosely packed regions, which correlate to various configurations and energy states in the potential energy landscape[6,7] (PEL), affect the properties (e.g. strength and plasticity) of MGs[8]. Thus, modulating energy states in the PEL through optimizing atomic configurations continue to be a challenge and current research area.

Generally, as cast MGs will naturally age or relax to lower energy states by annihilation of the free volume[9] or the flow units[10]. Contrary to aging effect, MGs can be rejuvenated to higher energy states by various mechanical[11-14] or thermal treatments[15,16] through introducing microstructural defects, which significantly modulate mechanical properties. The increased enthalpy in glasses could be regarded as a measure of rejuvenation and manifests as an exothermic heat of relaxation upon heating[17]. Other than mechanical and thermal treatments methods, high pressure (HP) is also an effective and controllable thermodynamic and kinetic way that can alter microstructure, energy states and properties of MGs. Much effort has been devoted to the pressure effect on properties of MGs, including glass transition[18], crystallization behavior[19] and phase transition[20]. In the previous studies, Jin et al. reported that structural relaxation to a lower energy state by free volume annihilation was observed after high pressure annealing (HP-annealing) [18]. It is also reported that when HP is applied at room temperature (RT), Ce-based MGs can undergo a reversible polyamorphic phase transition[21] or an irreversible transition from amorphous to single crystalline states at 25 GPa[22]. In this wok, HP-annealing on $La_{60}Ni_{15}Al_{25}$ MG was performed. The results show that the MGs are rejuvenated to higher energy states that can be preserved after release of pressure.

Bulk $La_{60}Ni_{15}Al_{25}$ rod samples were prepared by suction casting (see Supplementary). For all samples, initialization was performed using differential scanning calorimetry (DSC) to eliminate the difference in thermal history among the as cast samples (see Supplementary). After the initialization, the HP treatments were performed in a cubic anvil type large volume high-pressure apparatus as shown in Fig. S1. The samples were treated under two sets of HP conditions, one is keeping the same pressure (~5.5 GPa) while annealing at different temperatures; the other is keeping the same temperature (~482 K) while applying different pressures. As presented in Fig. S2, the samples treated under all conditions show that there are no remarkable crystalline diffraction peaks, indicating that no crystallization took place during the HP treatments.

When MGs are heated, the structure relaxation, which is represented by an exothermic peak below glass transition temperature $T_g$ in DSC curve, is accelerated and increases significantly when temperature approaches $T_g$. The total heat release during relaxation, represented by relaxation enthalpy $\Delta H_{rel}$, is an indicator of the energy state of glasses. The various treatments could decrease or increase energy of MGs, known as aging or rejuvenation, which can be characterized by positive or negative $\Delta H_{rel}$ in DSC measurements, respectively. Figure 1 (a) shows DSC curves of heat flow versus temperature for samples annealed from RT to 454 K (slightly below $T_g$) under 5.5 GPa for 1 hour. Glass transition temperature $T_g$, crystallization temperature $T_x$ and crystallization enthalpy are obtained to be 458 K, 523 K and -5.9kJ/mol, respectively. The crystallization enthalpy does not change upon the experimental conditions, indicating no crystallization occurs during HP treatments. When the samples are compressed under pressure at RT, no obvious change in relaxation enthalpy is observed [Fig. 1 (a), dashed line]. As the annealing temperature increases from RT to 454 K while maintaining the pressure at 5.5 GPa, the rejuvenation takes place and enhances with increasing temperatures, characterized by the enhanced exothermic peaks. The inset of Fig 1(a) illustrates the increased $\Delta H_{rel}$ after HP-annealing using the equation: $\Delta H_{rel} = \int_{T_0}^{T_1} \Delta C_p \, dT$, where, $T_0$, is the temperature before exothermic relaxation takes place and is set as 333.15 K, $T_1$ is a temperature in the supercooled liquid region and is set as 488.15 K. $\Delta C_p$ is the difference between the specific heats of the samples after initialization and HP-annealing. The corresponding relaxation enthalpy is presented in Fig. 1 (b). A relaxation enthalpy of 0.9 kJ/mol can be obtained for the sample treated at 454 K under 5.5 GPa, which is three times larger than that of the most rejuvenated samples treated by thermal cycling[15], and similar to those treated by plastic deformation[12, 13]. The rejuvenation induced by HP-annealing remained after keeping the sample at RT for two months, however part of it has been relaxed at RT (inset of Fig. 1 (b)). On the contrary, for the MG annealed at the same temperature under ambient pressure for 1 h, the thermal process in DSC is slightly endothermal (Fig. 1 (b) and Inset (a) of Fig. 2 (b)), which is consistent with the results of the previous studies[23]. The aging effect at the same temperature under ambient pressure suggests that HP plays a crucial role in the rejuvenation of MGs.

To study the pressure effect on rejuvenation, the MGs were annealed at the same temperature of 426 K for 1 hour under different pressures: ambient pressure, 1.0, 2.2, 4.0, 5.5 and 8.0 GPa. The corresponding DSC curves after HP-annealing are presented in Fig. 2 (a). When pressure is relatively lower (ambient pressure and 1.0 GPa), the MG sample annealed at 426 K aged to a lower energy state (blue dash line and cyan solid line in Fig. 2(a)). As the pressure rises to 2.2 GPa, an exothermic peak appears, and its intensity increases with increasing pressure and finally reaches a maximum at 8.0 GPa. The relaxation enthalpy (Fig. 2 (b)) increases with increasing pressure and reaches 0.9 kJ/mol at 8.0 GPa. The samples treated under different pressures at RT show that no rejuvenation occurs in DSC curves (Inset (b) of Fig. 2 (b)), suggesting that high temperature is also a necessary condition for the rejuvenation. Therefore, the rejuvenation of MGs is attributed to the coupling effect of high pressure and high temperature.

Figure 3 shows a schematic of the temperature dependent enthalpy change of glasses and supercooled liquids under different pressures. During sub-$T_g$ annealing, the enthalpy of MGs moves towards the equilibrium state, which is the extrapolated line of the supercooled liquid state. In the previous studies, when the pressure is relatively lower, the enthalpy of the supercooled liquid decreases slightly[18]. When the pressure increases to a certain extent, the enthalpy of supercool liquid will increase[24] (black dashed line in Fig. 3), which is confirmed by MD simulations (see Fig. S4). As a result, there is an intersection temperature, $T_{in}$, between the line of enthalpy for glass under ambient pressure (black solid line) and the line extrapolated from the supercooled liquid under HP (black dashed line). When temperature is higher than $T_{in}$, the enthalpy will increase towards (red arrow line) the HP equilibrium state upon HP-annealing. However, for ambient pressure aging still dominates the process and the enthalpy decreases (blue arrow line). If temperature is below $T_{in}$, the enthalpy will decrease towards the liquidus for HP (light blue dash arrow line), suggesting that both of high temperature and HP are necessary conditions for the rejuvenation. The rejuvenation can be preserved when the temperature and pressure are reduced to ambient conditions, yet the MG would slowly relax to the equilibrium state under ambient condition.

The rejuvenation induced by HP-annealing can also be interpreted by the potential energy landscape (PEL) theory[6,7]. The PEL contains many saddle points and local minima, which represent the potential energy states of MGs with various configurations. During sub-$T_g$ annealing, the MG has potential to jump across the energy barrier, fall into a local minimum with lower energy and finally relax to an ideal glass, after being annealed for sufficient time due to ergodicity[25]. However, HP could alter the PEL dramatically. Confirmed by previous studies, strain could induce disappearances of energy minima[26, 27]. Thus the system could age to a lower energy state after a small strain cycle [28]. However, when applied a loading-unloading cycle of large strain, the system rejuvenated to a higher energy state. In this case, the shape of the PEL has been vastly changed, and the system could undergo relaxations and locate in a local minimum far from the initial state, which signifies that the final configuration has changed greatly. After strain recovering, the preserved configuration may locate in the shallower minimum than that of initial configuration[28]. In the case of HP-annealing, the PEL changes under HP. As displayed in Inset (a) of Fig. 3, for a certain configuration, the energy state may locate in a basin under ambient pressure yet lie on a saddle point under HP. Accordingly, for the MG, which is annealed below $T_g$ under HP, the energy could be kept in a local minimum in the HP-PEL (from state 1 to 2). After lowering temperature to RT and then unloading the pressure, the configuration favored under HP is no longer preferred, but the temperature is too low to provide energy to overcome the energy barrier, and no notable atomic rearrangement could take place after unloading at RT. Consequently, the potential energy of the MG can only reside into local minima with a similar configuration to the one favored under HP (from energy state 2 to 3), which is not favored anymore under ambient pressure and hence has higher potential energy than that of the original initialized sample. Therefore, the higher energy state can be preserved after HP-annealing in the perspective of PEL theory.

Generally, high pressure promotes atomic rearrangement in short range. When pressure is relatively lower, free volume annihilation dominates the process and results in aging. Jin et al studied the pressure effect on structural relaxation in MGs based on the free volume model[18,29]. However, when the pressure is relatively higher, the MGs are further compressed and the atoms could rearrange into configurations with higher atomic packing density. As shown in Fig. 2 (b), the aging occurs under lower pressure (0.1 MPa and 1 GPa), but the rejuvenation emerges with increasing pressure. It could be understood by the Lennard-Jones potential (Fig.3 inset (b)), when the interatomic distance is larger or smaller than the critical distance (A) of the bottom of the potential well, the potential energy will increase. Under HP-annealing, the interatomic distance will first decrease towards the critical distance (from B to A), which leads to free volume annihilation that corresponds aging. When the pressure further increases, the interatomic distance will decrease to the left (from A to C) of the critical distance, which leads to a higher potential energy and a further densification of MGs.

To verify above assumption, we measured the density and elastic modulus of the rejuvenated MGs. Unlike previously reported results, where rejuvenation usually causes a decrease in density and elastic modulus[12,15], in the present study a reverse trend is observed. After annealing at 454 K under 5.5 GPa for 1 h, the density increases by 1.09 %, i.e. from 5.82 g/cm$^3$ to 5.88 g/cm$^3$. The elastic and shear moduli increase by 2.65 %, i.e. from 38.4 GPa to 39.4 GPa, and 3.15 %, i.e. from 14.3 GPa to 14.8 GPa, respectively. As previously reported, the rejuvenation in MGs originates from the generation of microstructural defects of the flow units[10], which usually act like "soft spots" with a lower atomic packing density and a lower coordination[30,31]. However, in the HP-annealing, the rejuvenation in enthalpy is accompanied by densification, implying a different microstructural origin and mechanism for the rejuvenation.

To investigate structural origin of the rejuvenation induced by HP-annealing, double aberration corrected scanning transmission electron microscopy (Cs-STEM) experiments were performed. Figure 4 (a, b) show the colored Cs-STEM dark-field images of the distribution of metallic atoms in the initialized and HP-annealed samples, respectively. Figure 4 (c, d) show the atomic resolution Cs-STEM images of the same area of Figs. 4 (a, b), in which projective random distribution of atoms (bright spots) is clearly visible after filtering. All images show amorphous structures and no crystallization occurs, which are consistent with XRD results. However, compared to the initialized sample, the HP-annealed sample shows much larger regions with higher atomic packing density (marked by the yellow spots in Fig 4. (b), which are separated by blue regions of the elastic matrix. In addition, in the Fast Fourier Transformation (FFT) images (inset of Fig. 4 (a, b)), the HP-annealed sample displays a halo ring slightly thinner and brighter than that of the initialized sample, implying an ascending trend of local structural ordering. This is also consistent with the secondary ring outside the halo ring where there is a sudden leap in intensity in the FFT image of HP-annealed sample. The intensity of the FFT images is integrated[32] (Fig. 4 (e)), and it has two distinct features, marked as *A* and *B*. For the HP-annealed sample, the *A* peak is narrower, indicating a trend to a higher short range ordering. While there are humps in the area *B*, implying the intensity at higher Q or the number of atoms with shorter interatomic distance increases. The larger amount of atoms with higher packing density make MGs in higher energy states, and this could be the structural origin of the rejuvenation.

The rejuvenation in MGs is usually accompanied by an increase in flow units with lower atomic packing densities (containing free volumes, Fig. 3, Inset (b)). However, the rejuvenation induced by HP-annealing originates from clusters with higher atomic packing density. They are termed as negative flow units, which possess higher density, higher energy due to stronger local stress inside. Opposite to normal flow units, negative flow units contain negative free volumes that has been confirmed by high-energy X-ray diffraction[33,34] and experimentally observed in colloidal system[35]. In the HP–annealing process, pressure first leads to free volume annihilation, resulting in structural relaxation to lower energy states and the increase in density, manifesting as aging. When the pressure increases to a certain value, the negative flow units could be produced accompanied by a further increase in density. According to the experimental observations from aging to rejuvenation, the flow units change from positive to zero, and finally to a negative value, simultaneously the density and modulus increase continuously.

In summary, the MGs can be rejuvenated to higher energy states by HP-annealing below $T_g$ due to the coupling effects of pressure and temperature. The observed rejuvenation is accompanied by an increase in density and modulus, due to the formation of the negative flow units, which have a higher atomic packing density. The difference between PELs under different pressures results in such unique rejuvenation. This study enriches the traditional theories of flow units with a lower atomic packing density, and shows experimentally the existence of negative flow units by HP treatments. Compared to conventional rejuvenation, the HP-induced rejuvenation demonstrates new features: higher density, modulus and higher energy with negative flow units.

**Acknowledgements**

The discussions with M. X. Pan, Y. Z. Li, T. P. Ge, T. F. Pei and R. J. Xue, and experimental assistance by D. Q. Zhao and L. Han are appreciated. The work was supported by NSF of China (Nos. 51271197, 51271195, 51522212 and 51421002), MOST 973 Program (No. 2015CB856800, 2012CB932704 and 2014CB921002) and the Strategic Priority Research Program of Chinese Academy of Sciences (Grant No. XDB07030200).

**Figure Captions**

Figure 1  Rejuvenation in $La_{60}Ni_{15}Al_{25}$ bulk MG after high pressure treatments under 5.5 GPa at different temperatures. **a,** DSC curves for different annealing temperatures under 5.5 GPa. Inset: The evaluation of relaxation enthalpy. **b**, Relaxation enthalpy for samples tested immediately after HP-annealing and after two months relaxation at RT. For comparison, relaxation enthalpy of annealed samples at ambient pressure is shown. The data points are for individual samples and the changes are larger than the standard deviation of ±0.02 kJ/mol (Error Analysis in Supplementary). The dashed lines are guide for eye. Inset: DSC curves for HP-annealed samples relaxed under ambient condition for two months.

Figure 2  Rejuvenation in $La_{60}Ni_{15}Al_{25}$ bulk MG after treatments at 426 K under different pressures. **a,** DSC curves for different pressures at 426 K. **b,** Relaxation enthalpy for samples treated at temperature (426 K) and RT (298 K) under different pressures. The dashed lines are the guide for the eye. Inset (a, b), DSC curves for samples annealed at different temperatures under under different pressures at RT, respectively.

Figure 3  Schematic of temperature dependent enthalpy change of glasses and supercooled liquids under different HP-annealing conditions. The rejuvenation or aging occurs in MGs upon HP-annealing below glass transition temperature. Inset: (a) PEL changes for MGs under different pressures upon HP-annealing. (b) Sketch of the Lennard-Jones potential, A, B, C schematically denote the average interatomic distance upon different HP-annealing conditions. c) The difference between "flow units" and "negative flow units".

Figure 4  Cs-STEM images of $La_{60}Ni_{15}Al_{25}$ samples. **a-b,** Colored Cs-high-angle

annular dark-field (HAADF) STEM images of the projective distribution of metallic atoms in both initialized and HP-annealed samples with 16.384 × 16.384 nm$^2$ scan range, respectively. Insets show the corresponding diffraction patterns of **a-b** obtained by FFT. **d-e,** Filtered atomic resolution Cs-STEM images of the same area of **a-b**. **e** Integrated intensity profile of FFT images in **a-b**. A and B denotes the differences between samples.

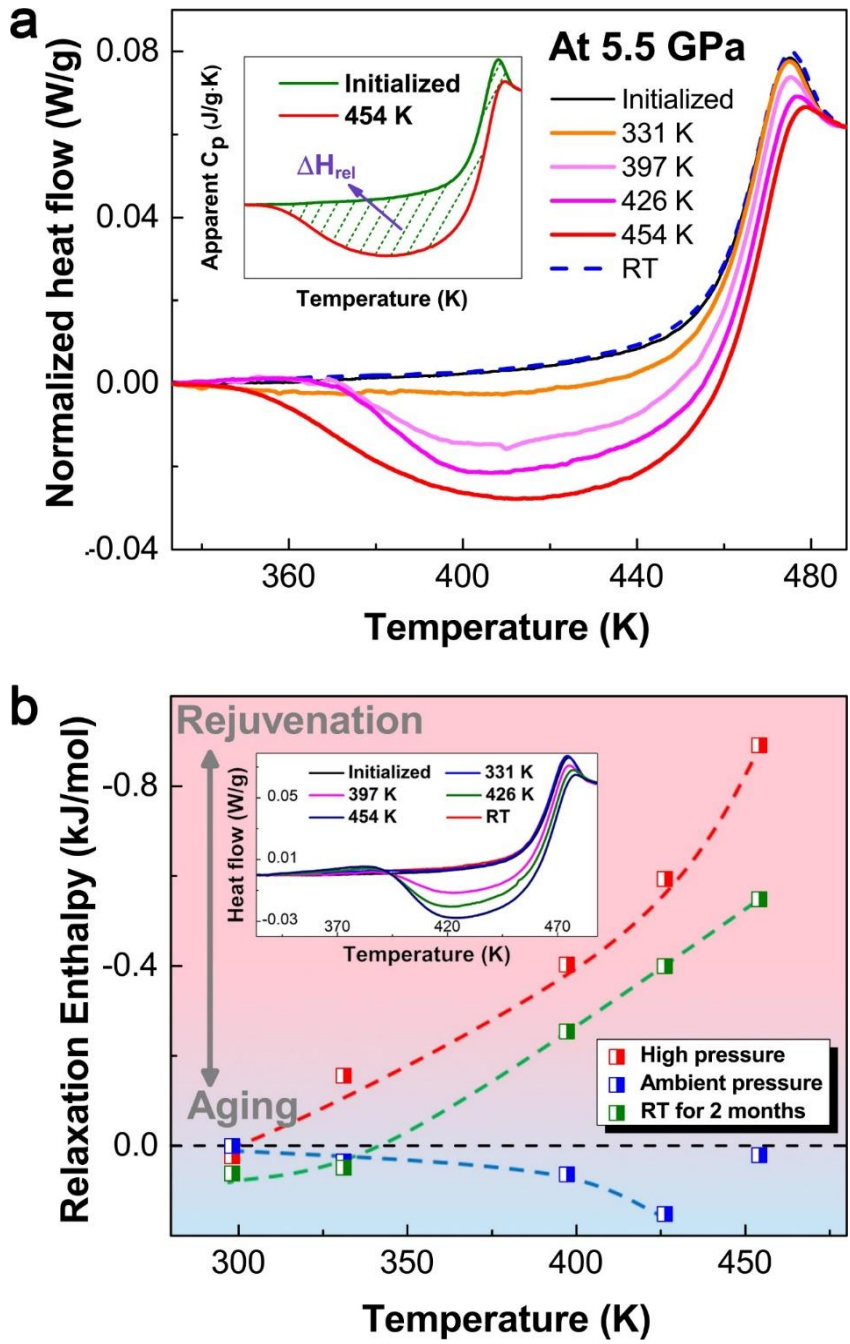

Figure 1. C. Wang, *et al*.

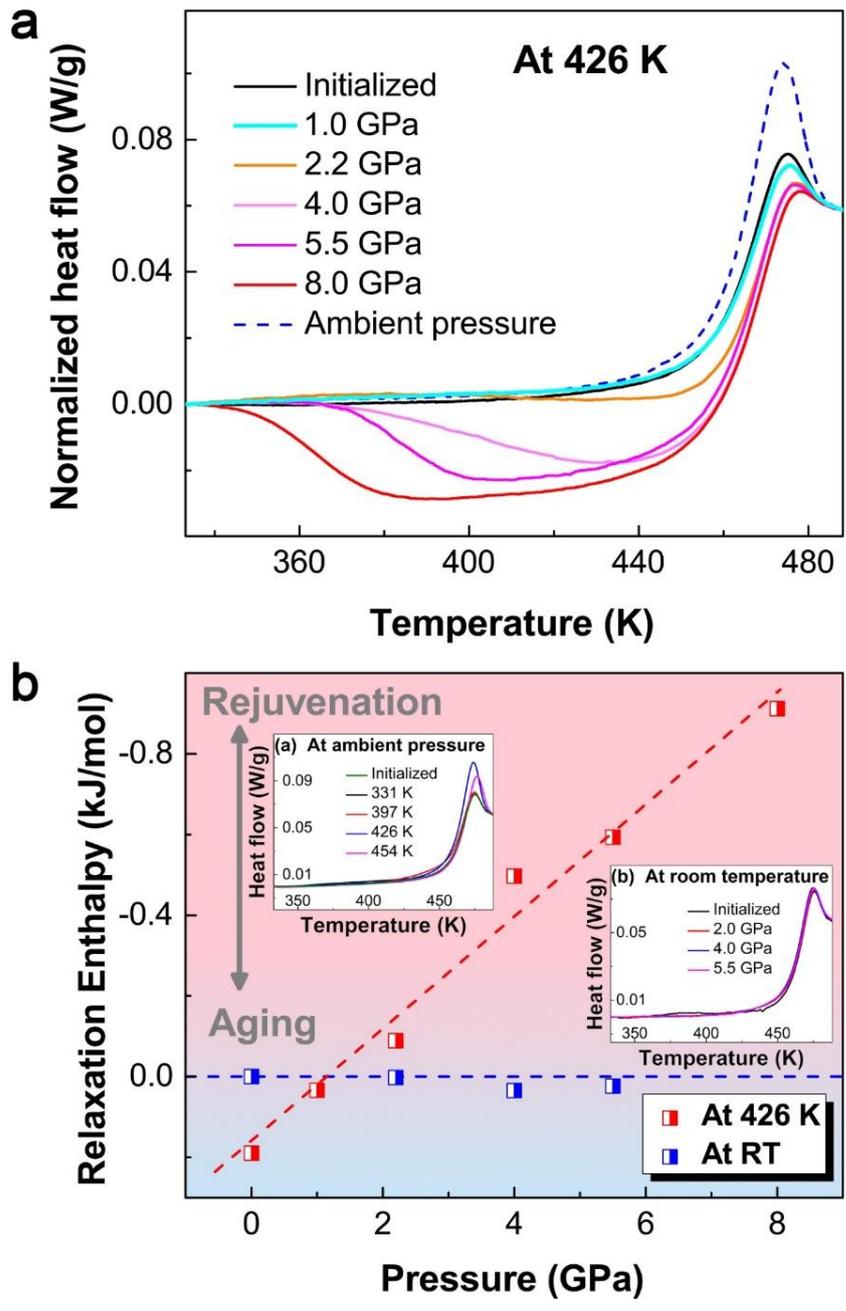

Figure 2. C. Wang, *et al*.

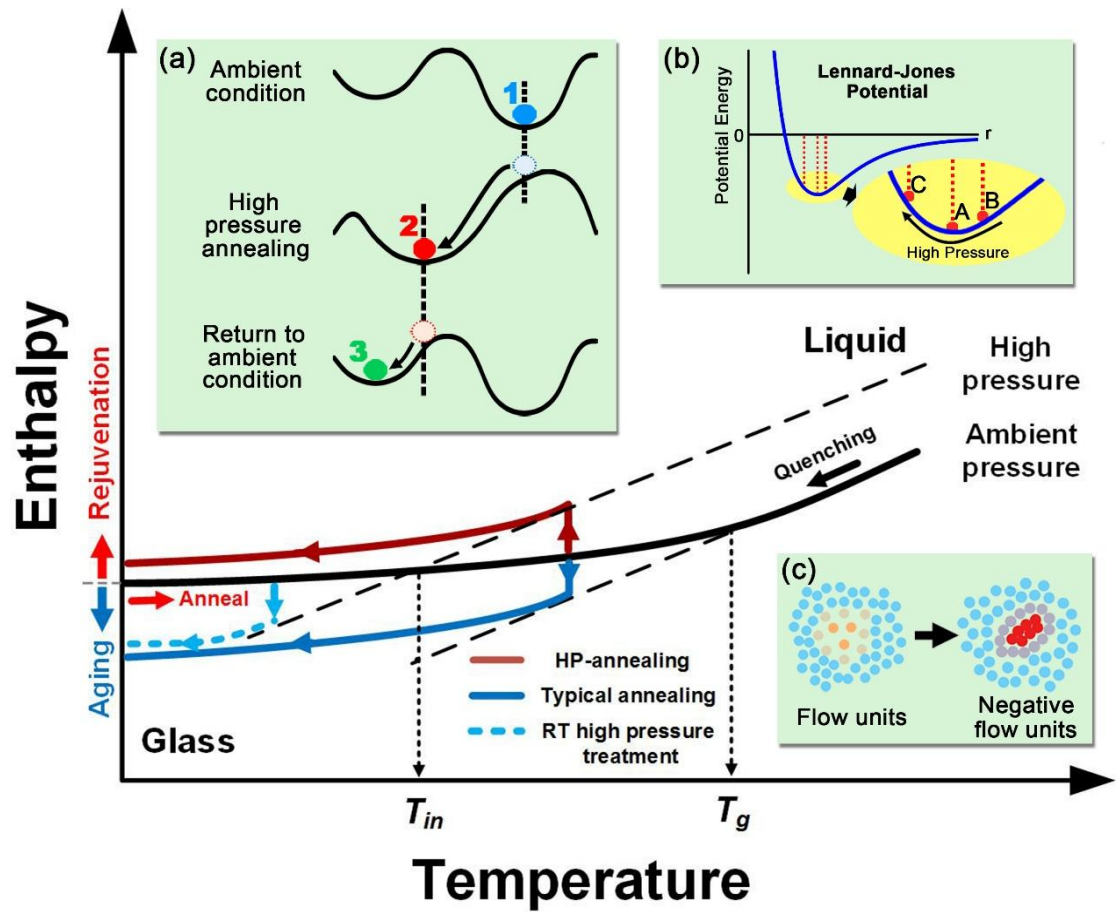

Figure 3. C. Wang, *et al.*

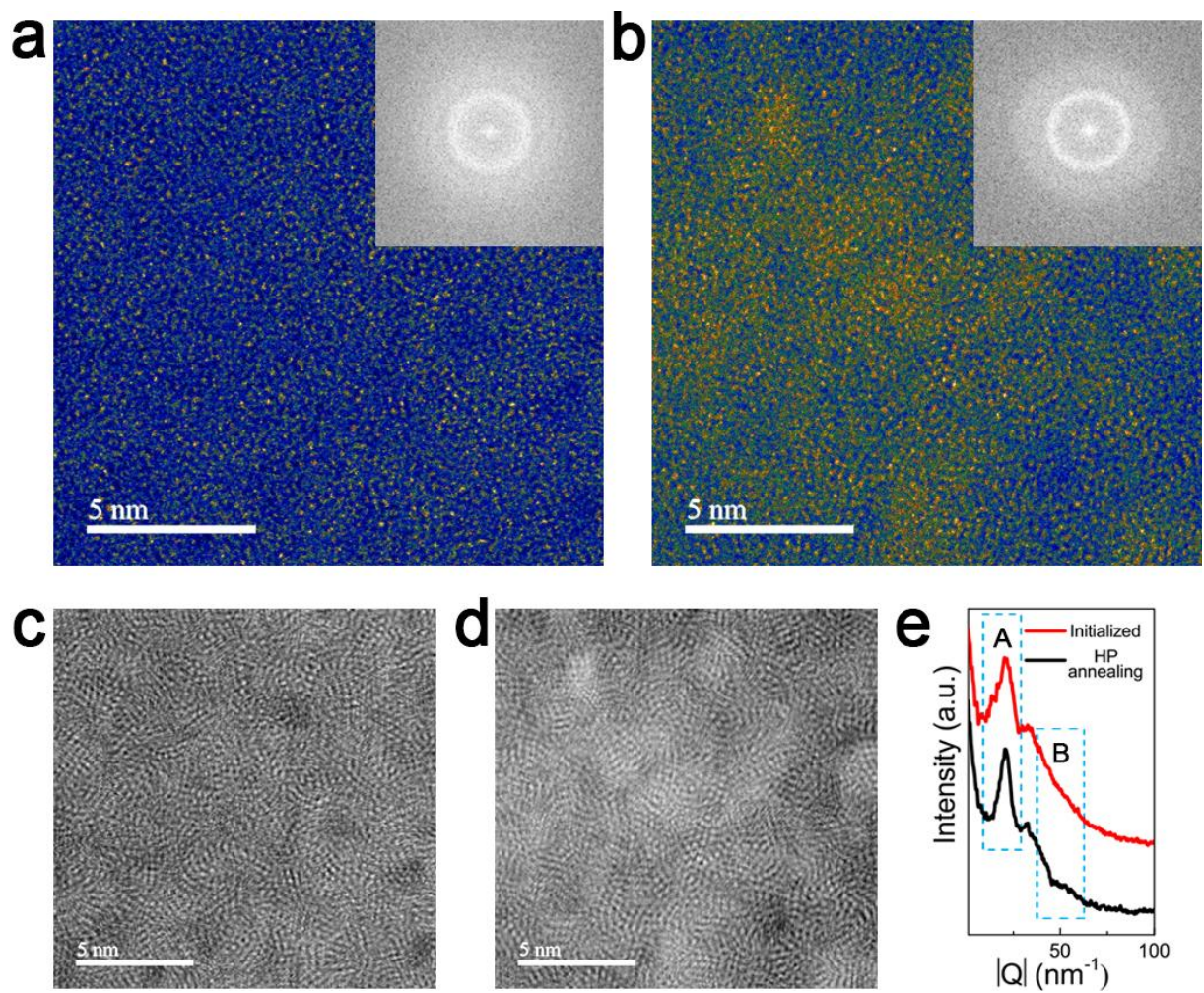

Figure 4. C. Wang, *et al.*

# Rejuvenation of metallic glasses under high pressure

C. Wang,[1] Z. Z. Yang,[1] T. Ma,[1,2] Y. T. Sun,[1] Y. Y. Yin,[1] Y. Gong,[1] L. Gu,[1,3] P. Wen, [2] P. W. Zhu,[1] Y. W. Long,[1] X. H. Yu,[1*] C. Q. Jin,[1] W. H. Wang,[1] H. Y. Bai[1*]

*[1] Beijing National Laboratory for Condensed Matter Physics and Institute of Physics, Chinese Academy of Sciences, Beijing 100190, China.*
*[2] State Key Laboratory of Superhard Materials, College of Physics, Jilin University , Changchun 130012, China.*
*[3] Collaborative Innovation Center of Quantum Matter, Beijing 100190, China.*

This file states the methods and details of sample preparation, DSC measurements, high pressure treatments, error analysis for relaxation enthalpy calculation, evolution of potential energy for glass forming liquids under different pressures, modulus and density measurements, and STEM sample preparation. It contains four supplementary figures to verify the claims made in the manuscript.

**Sample preparation**

Master alloy with a nominal composition of $La_{60}Ni_{15}Al_{25}$ (in atomic percentage) was prepared by arc melting pure elements under a Ti-gettered purified argon atmosphere. The master alloy was remelted at least five times to ensure composition homogeneity. Homologous rod-shape samples with diameters of 2 mm and 3 mm were prepared by suction casting master alloy into a water-cooled copper mold. The glassy nature was verified by XRD (Fig. S2) and DSC. Slices with a thickness of 1 mm were cut from the rod using a diamond saw. Both the upper and bottom surfaces were carefully grounded and polished using a designed grinding apparatus to ensure parallelism. The details of the bulk MG preparation can refer to as Ref [1].

**DSC measurements**

DSC experiments were performed using a power compensated Perkin-Elemer DSC 8000 under a high purity argon atmosphere. For the initialization, samples were heated up to 488 K into the supercooled liquid region with a heating rate of 0.33 K/s, without being held, and then cooled down to RT at the same rate of 0.33 K/s. After HP-annealing, the samples were heated from RT to 673 K, above the crystallization temperature, at 0.33 K/s, held for 2 minutes and then cooled to RT with the cooling rate of 1.66 K/s. The same procedure was performed for the crystallized sample for the baseline so that to subtract the baseline from the first cycle of heat treatment. At relatively lower temperatures before structural relaxation, no thermal behavior occurs, thus all DSC curves can be brought to intersect the *x* axis corresponding to zero heat flux at 333 K. Then the curves were corrected to accordant with each other in the supercooled liquid region with a linear correction. This linear correction has no influence on the features of exothermic process before occurrence of glass transition in DSC measurements. The relaxation enthalpy was calculated from the area between curves of the initialized sample and HP-annealed samples from the onset temperature of relaxation to a stable temperature in supercooled liquid region[2].

**HP treatments**

The HP treatments were performed in a cubic anvil type large volume high-pressure apparatus. For each HP experiment, the initialized $La_{60}Ni_{15}Al_{25}$ sample was placed in a cubic BN capsule, which was then sealed in a graphite heater. The detailed assembling is shown in Fig. S1. After calibration using a Ni-Cr/Ni-Si thermocouple, temperature was controlled by changing electrical power exerted on the graphite tube heater. At the target pressure ranges from 1-8 GPa, the temperature was gradually increased to 331, 397, 426 and 454 K respectively, and the pressure was kept for 1h before quenching to RT. After turning off the electric power, the sample temperature rapidly dropped to the ambient temperature because the anvils were surrounded by the cooling water tubes. Then, the sample was unloaded from HP conditions within 20 min for further characterizations.

**Error analysis for relaxation enthalpy calculation**

To assess the reproducibility and accuracy of the calculating relaxation enthalpy $\Delta H_{rel}$ using in the main text, four high pressure annealed samples under the same treatment condition were tested. Fig. S3 shows the corresponding DSC traces with an average value of $\Delta H_{rel} = 798$ J/mol, with a standard error of 22 J/mol (that is, less than ±3%). The relative change between samples under different conditions lies out this error range.

**Evolution of potential energy for glass forming liquids under different pressures**

In the Molecular Dynamics (MD) simulations, the potential energy of glass forming liquid under different pressures was investigated using LAMMPS package. The sample we used was a well-studied $Cu_{50}Zr_{50}$ system containing 10000 atoms, with embedded-atom method potentials. The samples were first equilibrated at 2000 K under a certain pressure (ranging from 0 GPa to 30 GPa). Then the samples were cooled down to a temperature below 1000 K with a cooling rate of 5000 K/ns and the pressure remained unchanged during this processes. All the simulation processes were done in NPT (constant number, pressure, temperature) ensemble. As shown in Fig. S4, the potential energy of each sample was monitored during the cooling process. As the pressure increases, the potential energy of the glass forming liquids first decreases when the pressure is small (less than 10 GPa), however increases when the pressure is further increased.

**Modulus and density measurements**

For modulus and density measurements, rods with diameter of 3mm were cut to a length of 6mm, and the upper and bottom surfaces were polished using a designed grinding apparatus to ensure parallelism with each other and perpendicular to the longitudinal axis. The acoustic longitudinal velocity and transverse velocity of the samples were measured using pulse echo overlap method at RT by MATEC 6600 ultrasonic system[3]. The frequency of the ultrasonic was 10 MHz and the sensitivity for measurement of time was 0.9 ns, which corresponded to a sensitivity of 5cm/s in velocity. The density was measured by Archimedes' principle in deionized water. The modulus and density changes were obtained from three individual samples which were measured before and after HP treatment, respectively. All samples show the increase in modulus and density after treatments and the evaluated mean values were listed in the main text.

**STEM sample preparation**

Considering that the sample is sensitive to mechanical force and high temperature, the sample were polished with a Twin-jet Electropolishing device to ensure the samples were suitable for STEM observation. During this process, the electrolyte was cooled by liquid nitrogen to temperature below -40 ℃. A JEM-ARM200F STEM fitted with a double aberration-corrector for both probe-forming and imaging lenses was used to perform HAADF imaging, which was operated at 200 KV. The convergence angle was 25 mrad and the angular range of collected electrons for HAADF imaging was about 70–250 mrad. All the scanning parameters are kept constant such as probe current and scanning speed. To better illustrate the difference between the initialized and HP-annealed samples, Fig. 4 (a, b) are colored under the same contrast.

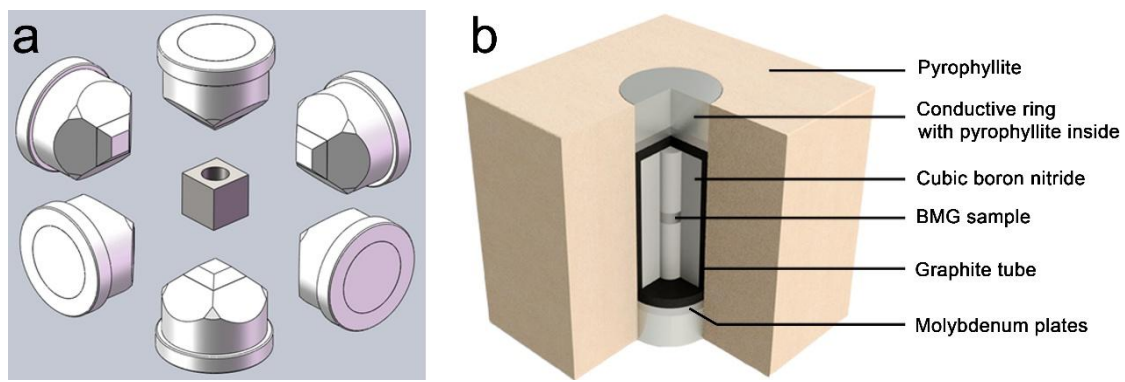

**Figure S1** A schematic of the cubic anvil type large volume high-pressure apparatus. a, the position of the pyrophyllite and six compressing tungsten carbide anvils. b, the assembling form of the BMG sample sealed in the pyrophyllite. The conductive ring is used for current conduction and contains pyrophyllite inside.

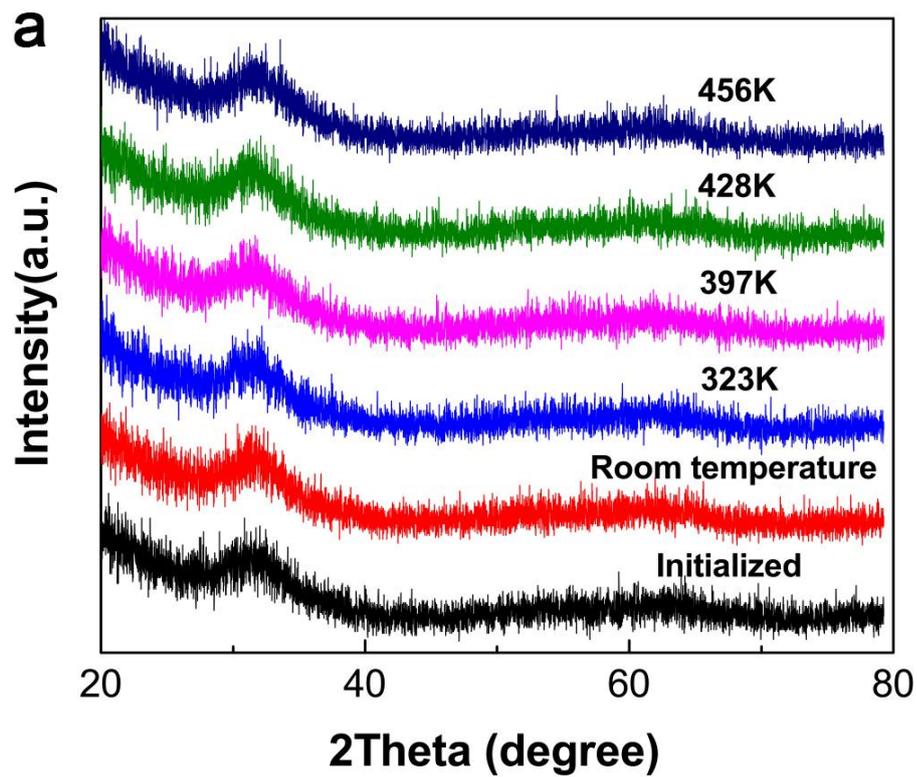

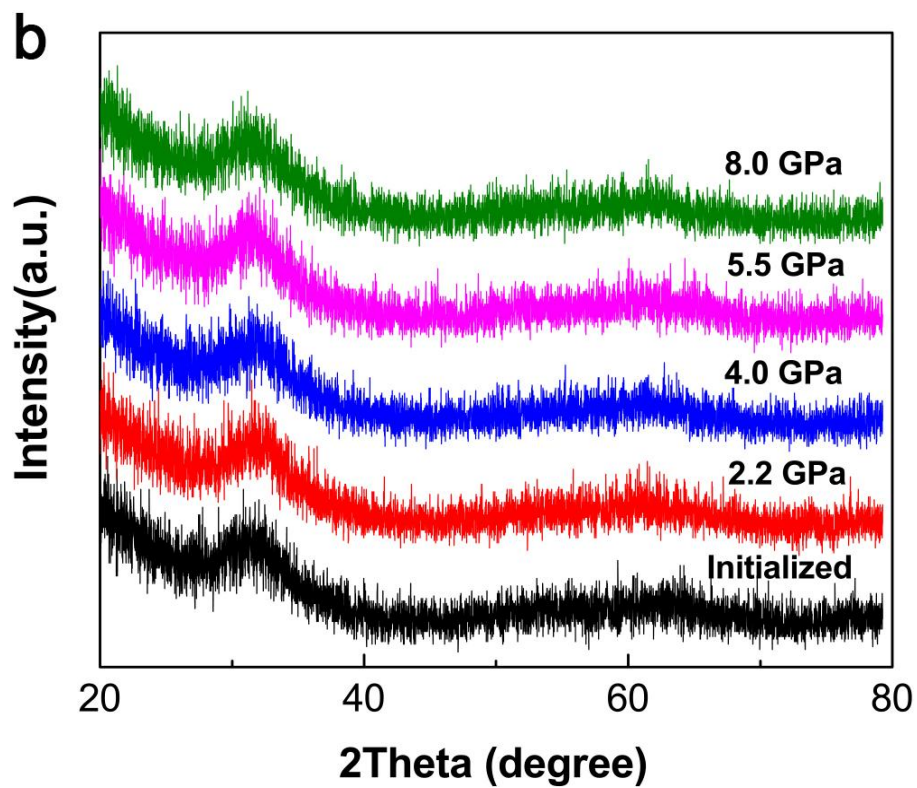

**Figure S2**  XRD diffraction patterns for samples treated at different temperatures under 5.5 GPa (a), and under different pressures at the same temperature of 426 K (b), respectively.

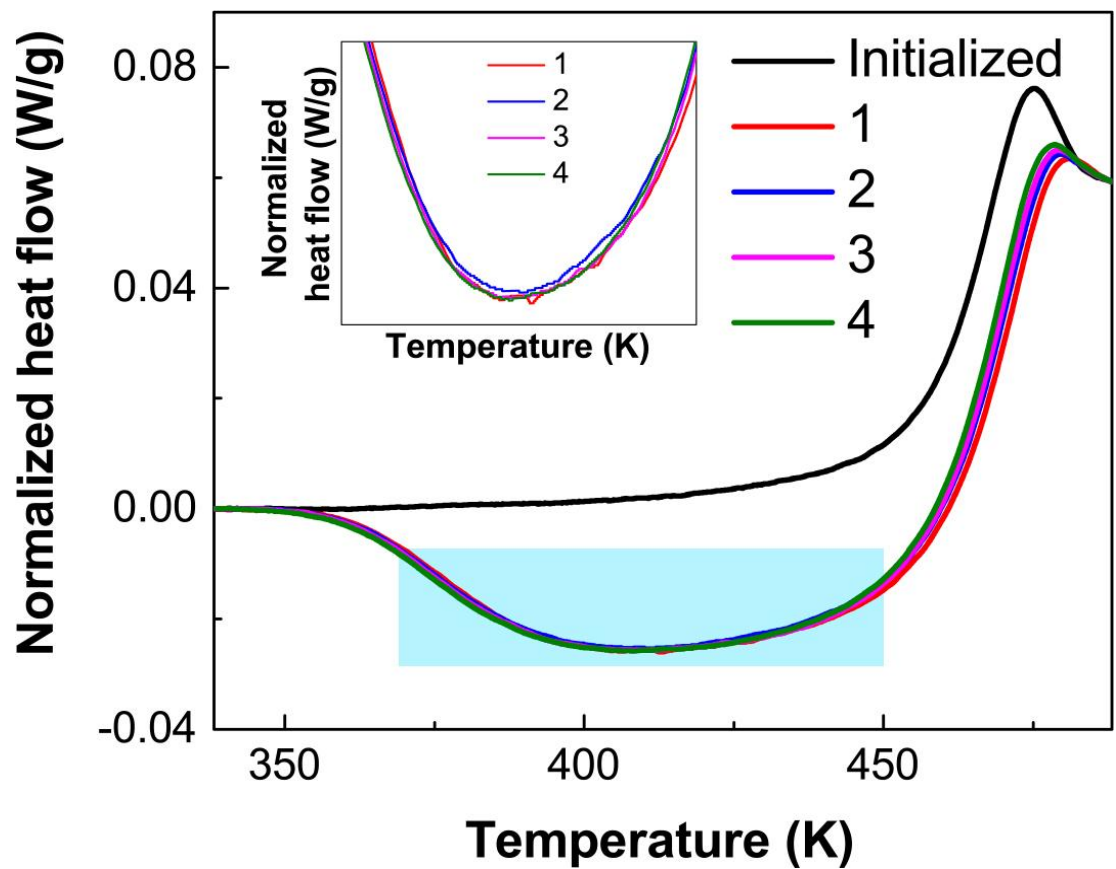

**Figure S3** DSC curves for the initialized sample and four HP-annealed samples under the same condition. Inset: enlargement of the light blue area showing the fluctuation of heat flow.

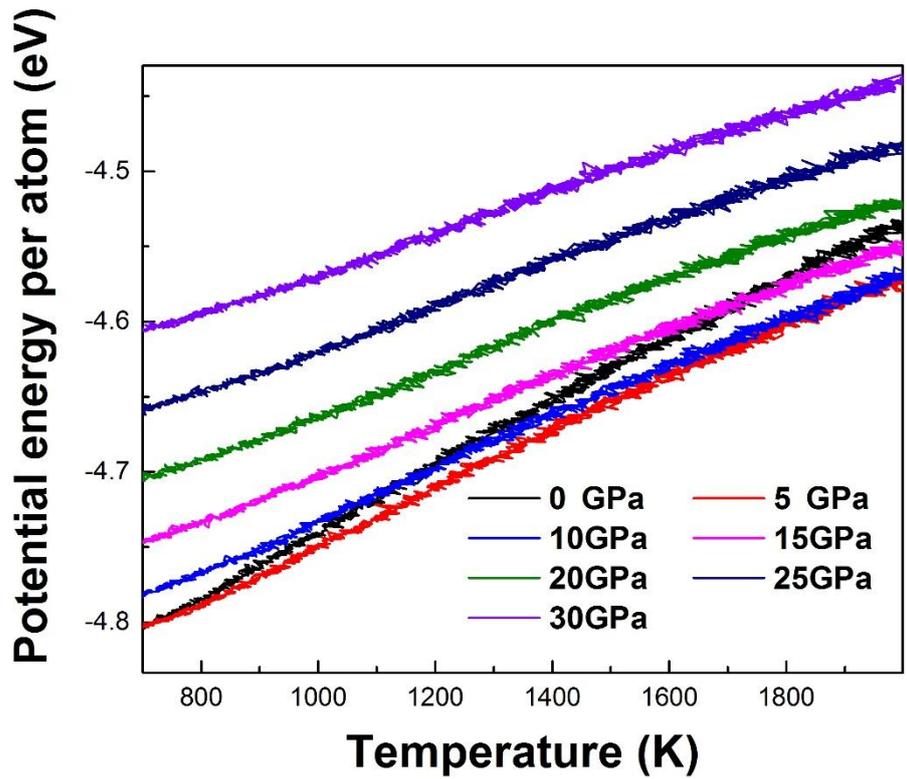

**Figure S4** The evolution of potential energy for glass forming liquids under different pressures.